\documentclass[aps,prl,groupedaddress,showpacs,notitlepage,twocolumn]{revtex4-1}
\usepackage{hyperref}
\usepackage{amsmath}
\usepackage{graphicx}
\usepackage{color}

\newcommand{\be}{\begin{equation}}
\newcommand{\ee}{\end{equation}}

\newcommand{\bs}{\boldsymbol}

\begin{document}

\title{Reconfigurable topological phases in next-nearest-neighbor coupled resonator lattices}

\author{Daniel Leykam$^{1}$, S.~Mittal$^{2,3}$, M.~Hafezi$^{2,3}$, and Y.~D.~Chong$^{4,5}$}

\affiliation{$^1$Center for Theoretical Physics of Complex Systems, Institute for Basic Science (IBS), Daejeon 34126, Republic of Korea\\
$^2$Joint Quantum Institute, NIST/University of Maryland, College Park MD 20742, USA \\
$^3$Department of Electrical and Computer Engineering and IREAP, University of Maryland, College Park MD 20742, USA  \\
$^4$Division of Physics and Applied Physics, School of Physical and Mathematical Sciences, Nanyang Technological University, Singapore 637371, Singapore \\
$^5$Centre for Disruptive Photonic Technologies, Nanyang Technological University, Singapore 637371, Singapore
}
\date{\today}

\begin{abstract}
We present a reconfigurable topological photonic system consisting of a 2D lattice of coupled ring resonators, with two sublattices of site rings coupled by link rings, which can be accurately described by a tight-binding model. Unlike previous coupled-ring topological models, the design is translationally invariant, similar to the Haldane model, and the nontrivial topology is a result of next-nearest couplings with non-zero staggered phases.  The system exhibits a topological phase transition between trivial and spin Chern insulator phases when the sublattices are frequency detuned. Such topological phase transitions can be easily induced by thermal or electro-optic modulators, or nonlinear cross phase modulation. We use this lattice to design reconfigurable topological waveguides, with potential applications in on-chip photon routing and switching.
\end{abstract}


\maketitle

\noindent

{\it Introduction.---} Topological photonic systems have attracted significant recent interest due to their potential applications as disorder-robust waveguides and delay lines~\cite{topological_photonics_review,review_2}. Topological protection is, however, both a blessing and a curse: steering the flow of light between different channels requires the ability to switch between topological phases by inducing band inversions. Such switching was recently demonstrated using mechanical reconfiguration of microwave photonic crystals and phononic metamaterials~\cite{ma2015,cheng2016,goryachev2016,susstrunk2017}. It would be interesting to achieve similar functionality in the optical domain, particularly for on-chip applications, using dynamic reconfiguration based on thermal, electro-optic, or nonlinear effects~\cite{lumer2013,ablowitz2014,nonlinear_SSH,leykam2016,kartashov2016,instability,hadad2017,NJP_paper,shalaev2017}. However, existing designs for topological photonic lattices at optical frequencies, such as helical waveguide arrays~\cite{rechtsman2013,leykam2016a,noh2017} or coupled resonators~\cite{hafezi2011,fang2012,hafezi2013,mittal2014,peano2015,minkov2016,mittal2016b} are ill-suited because they are based on spatial or temporal modulations and require inhomogeneous tuning to induce phase transitions. For example, the coupled resonator array studied in Ref.~\cite{hafezi2011} uses staggered phase shifts to simulate the integer quantum Hall effect, requiring individual control over these staggered phases.

Here we show that next-nearest neighbor hoppings provide a practical way to achieve reconfigurable topological phases in translationally invariant optical resonator lattices. Importantly, transitions between trivial and nontrivial phases are induced by resonance frequency shifts that are small compared to the rings' free spectral range. We demonstrate the reconfigurability of this system with two examples. In the first, thermal or electro-optic tuning~\cite{mittal2016b,electro_modulator} of the ring resonances reroutes edge states at an interface between trivial and non-trivial phases. In the second example, a strong pump induces nonlinear cross-phase modulation~\cite{nonlinear_review} which then triggers a topological phase transition. The features of our model thus provide a promising route to achieving reconfigurable topological edge modes in optical devices, as well as for studying how optical nonlinearities affect photonic topological edge states.

We consider a 2D bipartite lattice consisting of resonant ``site rings'' coupled via off-resonant ``link rings,'' with the latter positioned such that they facilitate both nearest-neighbour (NN) and next-nearest neighbor (NNN) hoppings between the sites. The two circulations of light (clockwise/anticlockwise) in each sublattice form a pseudo-spin degree of freedom; within each spin sector time-reversal ($\mathcal{T}$) symmetry is effectively broken, enabling quantum spin Hall edge states that are immune to backscattering as long as the spins are decoupled~\cite{hafezi2011}. Similar to the Haldane model~\cite{haldane_model}, the simultaneous presence of next-nearest neighbor hoppings and $\mathcal{T}$-breaking gives rise to a phase diagram hosting both trivial (conventional insulator) and nontrivial (spin Chern insulator) phases. By contrast, the nearest neighbor-coupled design introduced by Hafezi et al.~\cite{hafezi2011} broke translational symmetry by assigning uniformly increasing phase shifts to the link rings, resulting in a fractal Hofstadter butterfly spectrum~\cite{hafezi2011,hafezi2013,mittal2014,mittal2016b} and lacking simple topological transitions based on band inversions. On the other hand, band inversions in ring resonator lattices \textit{without} aperiodic elements can also occur in the strong coupling limit described by scattering matrices~\cite{liang2013,pasek2014,hu2015,spoof_plasmon,shi2017}, but strong coupling implies low quality factor resonators, which is not useful for the enhancement of nonlinear effects or delay lines.

\begin{figure}
  \includegraphics[width=\columnwidth]{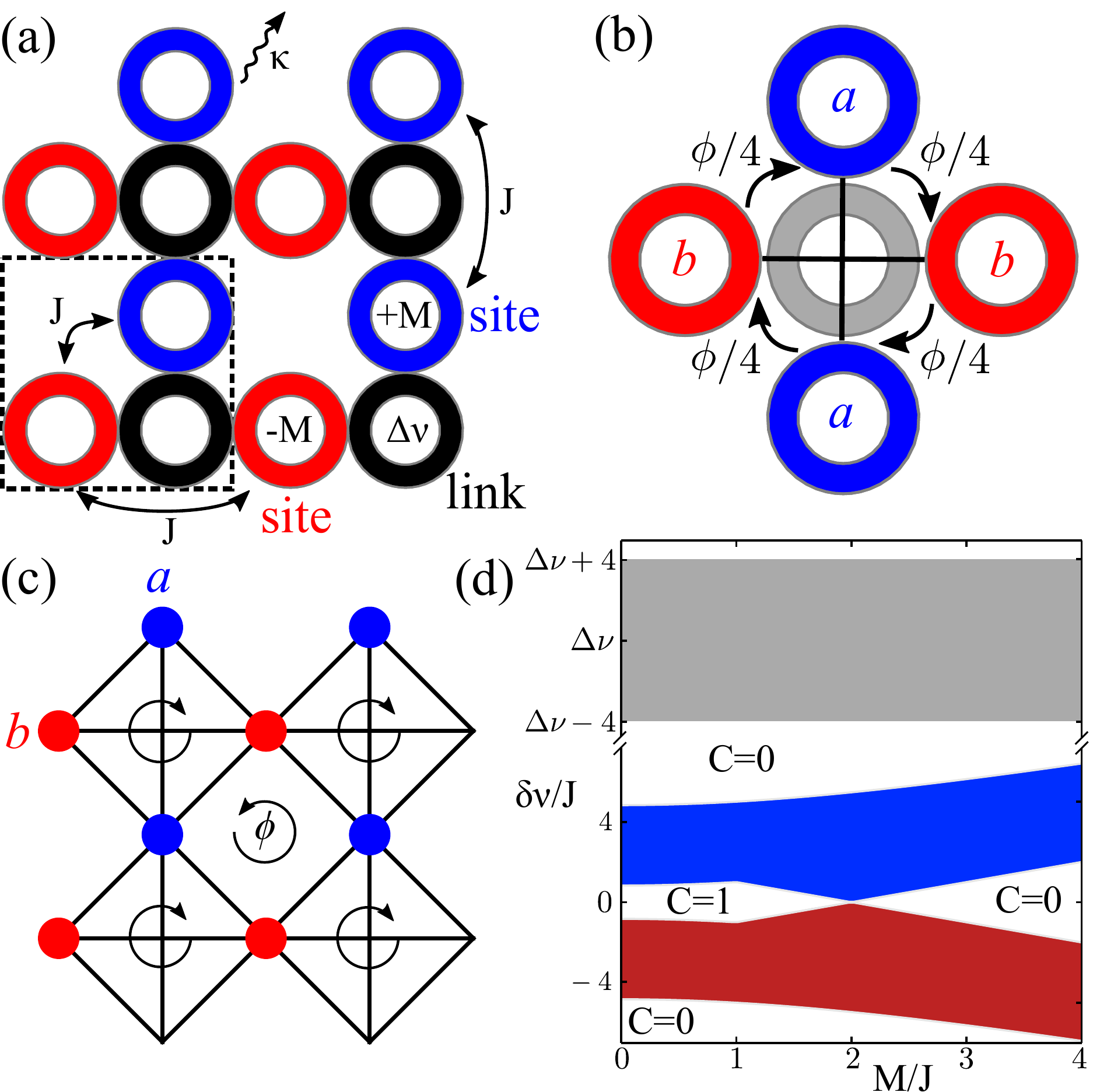}
  \caption{(a) Square resonator lattice of site rings (blue, red) with relative detuning $\pm M$, coupled with strength $J$ via off-resonant link rings (black) detuned by $\Delta \nu$. All rings have intrinsic loss $\kappa$. Dashed line indicates the unit cell. (b) Detailed schematic centered on a single link ring, which mediates both nearest-neighbor and next-nearest-neighbor couplings between site rings, with coupling phases $\pm \phi/4$ and 0 respectively. (c) Tight binding model for the site ring resonances, consisting of a checkerboard lattice  with staggered flux $\pm \phi = \pm 2 \pi \Delta \nu / \mathrm{FSR}$.  (d) Bloch band energies $\delta \nu$ (shaded regions) versus site detuning $M$, calculated from the tight-binding model for anti-resonant site and link rings ($\phi = \pi$). Red/blue areas indicate the sublattice the modes are localized to in the large $M$ limit. Gap Chern numbers $C$ are indicated and topologically nontrivial link bands occur for $|M|< 2J \csc \phi/2$.}
\label{fig:lattice}
\end{figure}

Figure~\ref{fig:lattice}(a) shows a schematic of the 2D lattice. The site rings (shaded red and blue) are positioned on a checkerboard lattice, while the link rings (shaded black) are positioned such that they couple the nearest neighbor site rings (diagonal couplings between red and blue sites) as well as the next-nearest neighbor site rings (horizontal and vertical couplings between red and blue sites, respectively). Furthermore, as shown in Fig.~\ref{fig:lattice}(b), the nearest-neighbor couplings between site rings (diagonals) introduce a direction-dependent hopping phase, which arises from the difference in the path lengths traveled in the link rings while hopping in different directions. The NNN hoppings are symmetric and do not carry a direction-dependent phase. In the absence of the NNN couplings, the system is gapless with a pair of Dirac points; the NNN couplings contribute to opening a nontrivial band gap~\cite{haldane_model,NNN_hopping}. Note that unlike the system of Ref.~\cite{hafezi2011}, the NN hopping phases in our system are periodic, and therefore the lattice is translation invariant.

{\it Tight binding model.---} We derive a tight binding model for the lattice by restricting ourselves to one of the two decoupled circulation sectors (anticlockwise site modes)~\cite{hafezi2011}, and taking the operating frequency to be close to the resonance frequency of the site rings, which are detuned from the link rings by $\Delta \nu$. The effective inter-site coupling can be derived by considering two site rings connected by a link ring, as in Ref.~\cite{hafezi2013}, and depends on the parameter 
\begin{equation}
  \phi = 2\pi\,\Delta\nu\,/\,\mathrm{FSR},
\end{equation}
where $\mathrm{FSR} \approx 10^3$ GHz is the rings' free spectral range~\cite{hafezi2013} (we assume the site and link rings have similar dimensions).

We assume that the rings are weakly-coupled high quality factor resonators, so that $J \ll (\Delta\nu,\,\textrm{FSR})$, where $J$ is the inter-site coupling strength for antiresonant sites and links (i.e., when $\phi = \pi$). Moreover, we apply a small detuning $M$ between the two site ring sublattices; inversion symmetry is broken for $M \ne 0$, which will be the mechanism for inducing a topological transition.  Using the weak-coupling approximation to eliminate the link ring amplitudes as modal variables, we derive the tight binding Hamiltonian $\hat{H}$~\cite{supplementary}:
\begin{align}
\hat{H} &= 
\sum_{x,y} \left( \hat{H}_a + \hat{H}_b + \hat{H}_{ab} + \hat{H}_{ab}^{\dagger} \right), \\
\hat{H}_a &= \hat{a}_{x,y}^{\dagger} \left[ (2J \cot \frac{\phi}{2} + M) \hat{a}_{x,y} + J \csc \frac{\phi}{2} \sum_{\pm} \hat{a}_{x,y\pm1} \right], \nonumber \\
\hat{H}_b &= \hat{b}_{x,y}^{\dagger} \left[ (2J \cot \frac{\phi}{2} - M) \hat{b}_{x,y} + J \csc \frac{\phi}{2} \sum_\pm \hat{b}_{x\pm1,y}\; \right], \nonumber \\
\hat{H}_{ab} &= J e^{i \phi/4} \csc \frac{\phi}{2} \nonumber\\ & \;\times\Bigg[ \hat{a}_{x,y}^{\dagger} (\hat{b}_{x,y} + \hat{b}_{x+1,y+1} )
  + \hat{b}_{x,y}^{\dagger} (\hat{a}_{x-1,y} + \hat{a}_{x,y-1} ) \Bigg]. \nonumber
\end{align}
Here, $\hat{a}^{\dagger}$ and $\hat{b}^{\dagger}$ are creation operators for the $a$ and $b$ sublattices respectively, $(x,y)$ are integers indexing the lattice sites, the effective coupling is $J \csc \frac{\phi}{2}$, and $J \cot \frac{\phi}{2}$ is a coupling-induced frequency shift.

Note that the NN and NNN coupling strengths are equal in magnitude. By contrast, in the topological photonic lattices studied in Refs.~\cite{rechtsman2013,leykam2016a,noh2017,hafezi2011,fang2012,hafezi2013,mittal2014,mittal2016b}, NNN couplings are either absent or negligible compared to NN couplings. In momentum space, the Schr\"odinger equation governing the evolution of the field amplitudes $\psi = (\psi^{(a)},\psi^{(b)})$ can be compactly written as
\begin{align}
i \partial_t \psi &= \Big[\delta \nu - \hat{H}(k_x,k_y) \Big] \psi, \label{eq:link_model} \\
\hat{H} &= J \csc(\phi/2)\, \left( \begin{array}{cc} d_0 + d_z & d_x - i d_y \\ d_x + i d_y & d_0 - d_z \end{array} \right), \nonumber \\
d_0 &=  2 \cos(\phi/2) + \cos k_x + \cos k_y,  \nonumber \\
d_x &= 4 \cos(\phi/4) \, \cos(k_x/2)\, \cos(k_y/2), \nonumber \\
d_y &= -4 \sin(\phi/4)\, \sin(k_x/2)\, \sin(k_y/2), \nonumber \\
d_z &= \frac{M}{J} \sin(\phi/2) -\cos k_x + \cos k_y. \nonumber
\end{align}
Here, $\delta \nu \equiv \nu - \nu_0 - i \kappa$, where $\nu$ is the operating frequency, $\nu_0$ is the site resonance frequency, and $\kappa \ll \textrm{FSR}$ is the loss rate in each ring.  

If the two types of site rings are identical ($M = 0$), the two bands of $\hat{H}$ have Chern numbers $C=\pm1$, with a gap of size $\Delta = 2 J \sec^2 (\phi / 8)/[1 + \tan(\phi / 8)] \sim 2J$, containing topologically protected edge states.  As shown in Fig.~\ref{fig:lattice}(d), we can induce a topological transition into a trivial phase (where the bands' Chern numbers are zero) by varying $M$.  The transition point lies at $M = 2J \csc(\phi/2) \sim J \ll (\Delta\nu,\,\textrm{FSR})$.  This implies that the transition can be realized via weak physical effects, affecting either the site rings (varying $M$) or the link rings (varying $\Delta\nu$ and hence $M_c$).  At the transition point, the band structure contains a single Dirac point at either $(k_x,k_y) = (0,\pi)$ (if $M>0$) or $(\pi,0)$ (if $M<0$).

We have tested the validity of the tight binding model by comparing the bulk spectrum of $\hat{H}$ to the transfer matrix description.  The spectra are in good agreement, with band edge frequencies accurate to within 10\% for the moderate coupling strengths typically used in experiment ($J/\mathrm{FSR} \sim 0.01$). For large couplings ($J/\mathrm{FSR} \approx 0.06$ for $\phi = \pi$), the two-band tight binding approximation breaks down, in which case either a three-band tight binding model or the full transfer matrix formalism must be used.  Details are given in the Supplementary Material~\cite{supplementary}.

Having derived an accurate tight-binding Hamiltonian, we can use it in schemes for routing topological edge states~\cite{ma2015,cheng2016,goryachev2016,susstrunk2017}, manipulating topological edge states with optical nonlinearities~\cite{nonlinear_SSH,leykam2016,hadad2017,NJP_paper}, and other interesting possibilities.  Two examples are presented below.

{\it Reconfigurable domain wall.---}The resonance frequencies of optical ring resonators can be actively controlled using on-chip thermal~\cite{mittal2016b} or electro-optic modulators~\cite{electro_modulator}.  By using such methods to adjust the detuning $M$ between vertical and horizontal site rings, we can selectively induce a topological transition in part of a lattice (without any spectral shift), producing domains with topologically distinct gaps at the same frequency.  This would allow us to realize a reconfigurable topological waveguide~\cite{ma2015,cheng2016,goryachev2016,susstrunk2017}.

\begin{figure}[t]
  \includegraphics[width=\columnwidth]{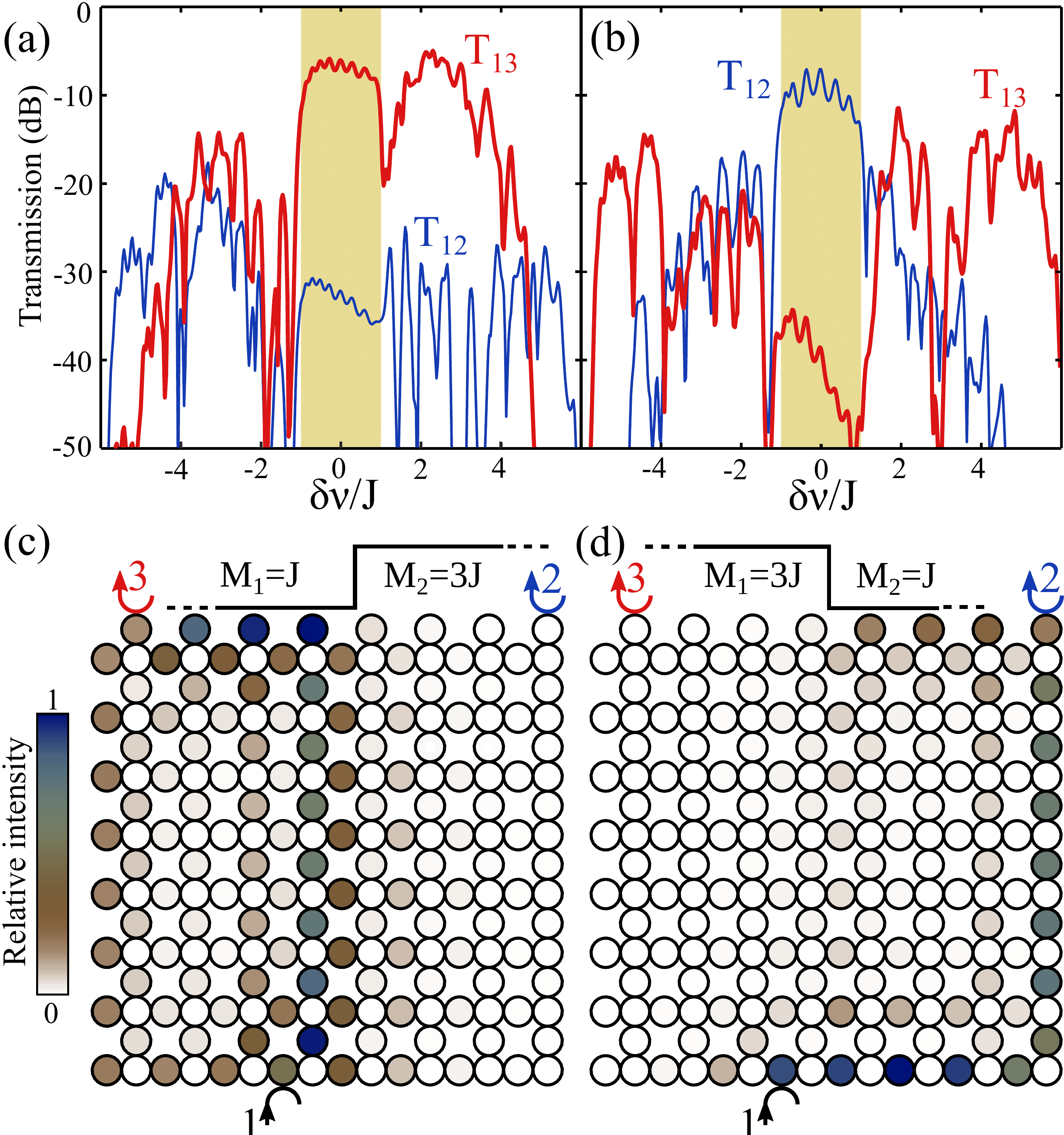}
  \caption{Reconfigurable topological waveguide functionality enabled by active control of site ring detunings.  The lattice contains $8 \times 8$ unit cells, with two domains where $M = M_1$ and $M = M_2$ respectively, with $\kappa = 0.05J$ and $\phi = \pi$ (i.e., site and link rings detuned by $\mathrm{FSR}/2$) throughout.  The couplings at the inputs and outputs are $\kappa_{\mathrm{ex}} = 2J$. (a) Transmission curves for $M_1 = J$ and $M_2=3J$; the positions of the input port ($1$) and output ports ($2,3$) are shown in (c), and the band gaps are shaded in yellow.  (b) Transmission curves for $M_1 = 3J$ and $M_2=J$.  (c)--(d) Intensity profiles at the mid-gap frequency $\delta\nu = 0$.}
  \label{fig2}
\end{figure}

Fig.~\ref{fig2} shows an example of such a scheme.  The lattice contains two domains, with $M = M_1$ on the left and $M = M_2$ on the right, with all other lattice parameters the same.  By controlling these detunings, we can define a domain with a nontrivial gap (with $M = J$) and a domain with a trivial gap (with $M = 3J$).  In practice, this can be accomplished by fabricating an array with uniform site offsets $\nu_a = J$ and $\nu_b = -3J$, and then in each domain detuning either $\nu_a$ or $\nu_b$ by $2J$.  In other words, we need only a binary shift of the site resonances (0 or $2J$), rather than the inhomogeneous shifts required in Ref.~\cite{hafezi2013}, or the much stronger shifts ($\approx \mathrm{FSR}/2$) in the strongly-coupled ring lattice discussed in Ref.~\cite{spoof_plasmon}.

As shown in Fig.~\ref{fig2}(c)--(d), we define an input port 1 that couples to a single site ring, and output ports 2 and 3 on opposite corners of the lattice.  The input and output couplings are described by
\begin{align}
  i\partial_t \psi^{(b)}_{\mathrm{in}} &= (\delta \nu - i \kappa_{\mathrm{ex}}-\hat{H}) \psi^{(b)}_{\mathrm{in}} + i\sqrt{2\kappa_{\mathrm{ex}}} \varepsilon_{\mathrm{in}}, \\
  i\partial_t \psi^{(a)}_{\mathrm{out,j}} &= (\delta \nu - i \kappa_{\mathrm{ex}}-\hat{H}) \psi^{(a)}_{\mathrm{out,j}},
\end{align}
where $\kappa_{\mathrm{ex}}$ is the input/output coupling rate, $\psi^{(b)}_{\mathrm{in}}$ denotes the input site, and $\psi^{(a)}_{\mathrm{out},j}$ ($j = 1,2$) are the output sites.  From these, we can compute the steady-state transmittances $T_{1j} = 2 \kappa_{\mathrm{ex}} |\psi^{(a)}_{\mathrm{out,j}}|^2/\varepsilon_{\mathrm{in}}^2$. 

Fig.~\ref{fig2} plots the transmission spectra and mid-gap intensity profiles for the two choices of interface orientation, with $\kappa = 0.05J$, $\kappa_{\mathrm{ex}} = 2J$, and lattice size $N=8$~\cite{hafezi2013}. Broad transmission maxima limited only by the intrinsic absorption $\kappa$ occur in the band gap of the array, mediated by topological edge modes, with the transmission to the alternate output port suppressed by over 30dB. In the Supplementary Material, we show that this transmission is robust against spin-conserving disorder~\cite{supplementary}, which is the most significant source of backscattering in typical coupled resonator optical waveguides~\cite{hafezi2013,mittal2014}. Moreover, the large topological band gap can approach the free spectral range in experimentally-realistic settings.  For example, in Ref.~\cite{mittal2016b} the thermo-optic modulators induced shifts of up to $0.3\,\mathrm{FSR} = 300$ GHz. Employing a similar shift in this design allows for reconfigurable topologically-protected pass bands of width $200$--$300$ GHz, which is promising for wavelength-division multiplexing applications.

{\it Pump-induced topological transition.---}The tight-binding model can also be used to design a device exhibiting nonlinearity-controlled topological transport. To demonstrate this, we fix the detuning $M$ and use cross phase modulation (acting on $\phi$) to tune between trivial and nontrivial phases. A strong pump, resonant with the link rings, shifts their resonance frequency such that~\cite{agarwal_book}:
\be 
\phi = \phi_0 - 4 \pi \nu_{\mathrm{NL}} |\psi^{(s)}_{x,y}|^2/\mathrm{FSR}, \label{eq:link_phase}
\ee
where $\phi_0$ is the coupling parameter in the absence of the pump, $\nu_{NL}$ is the effective Kerr coefficient, and the pump beam profile $\psi^{(s)}_{x,y}$ is governed by a square lattice tight binding model for the link rings~\cite{supplementary}. Assuming a uniform pump intensity $|\psi^{(s)}_{x,y}|^2 = I$, one can calculate the band structure and Chern numbers for a weak probe beam using the linear tight binding model Eq.~\eqref{eq:link_model} with effective coupling parameter Eq.~\eqref{eq:link_phase}; a phase transition between trivial and nontrivial phases occurs at the critical pump intensity $I = \mathrm{FSR}[\phi_0-\sin^{-1}(2J/M)]/4\pi\nu_{NL}$.

\begin{figure}[t]

\includegraphics[width=\columnwidth]{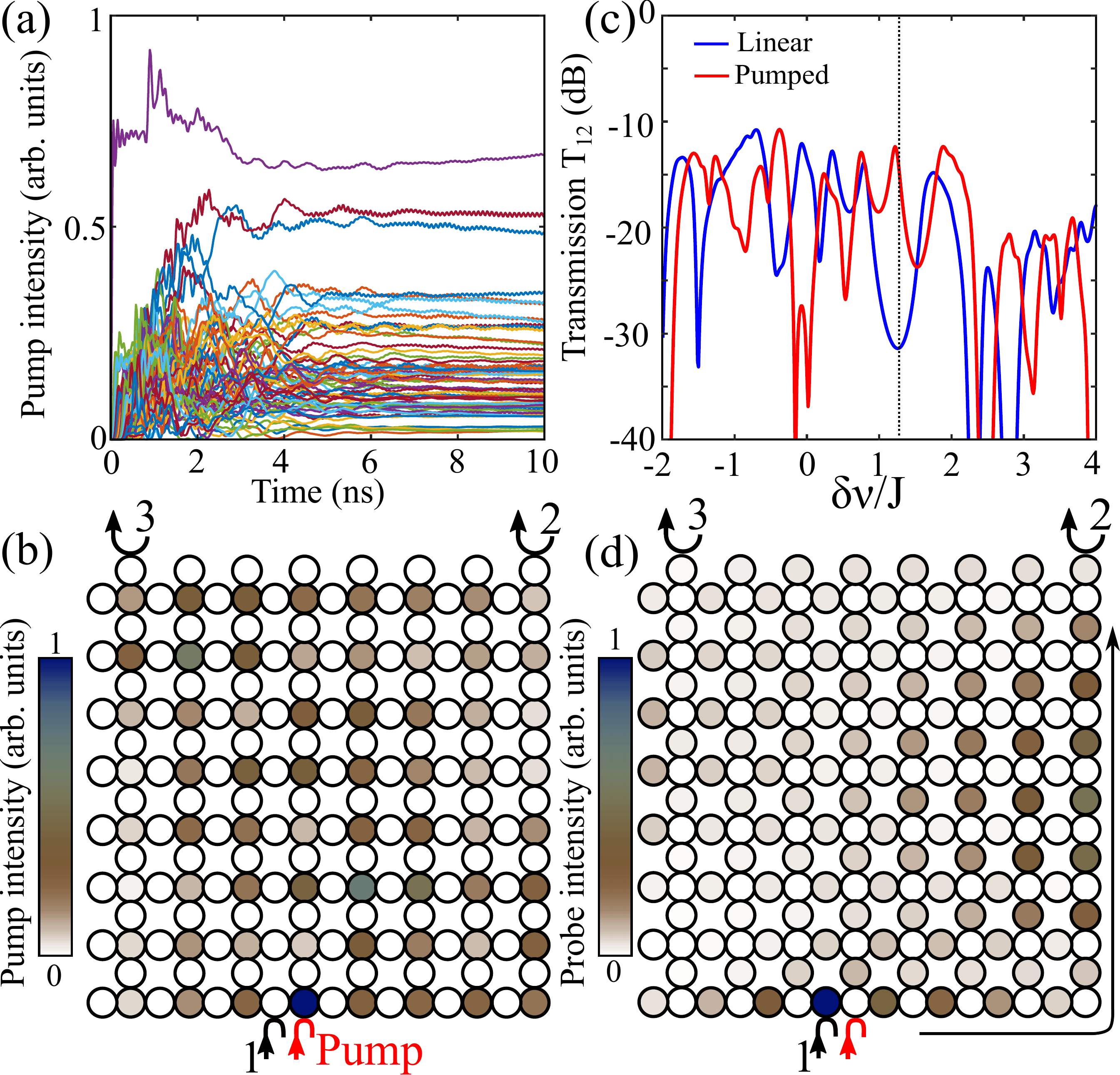}

\caption{Nonlinearity-induced topological transition. (a,b) Solution of the nonlinear scattering equations for a pump beam at $\nu - \nu_L = 4.6J$. (a) Dynamics of link ring intensities, illustrating relaxation to a stable steady state within 5ns. The different colors denote different link rings. (b) Resulting pump intensity profile over the $8\times 8$ lattice inducing a Chern insulator phase at the site resonance. Red arrow indicates the pump input position relative to the probe ports (1,2,3). (c) Transmission spectrum at the site resonance without (blue) and with (red) the pump beam. (d) Mid gap probe profile when pump is applied [frequency indicated by dashed line in (c)].}

\label{fig:pump_profile}

\end{figure}

As an example, we consider a homogeneous lattice with $M=-2.4J$ and $\phi_0 = 0.65\pi$ (in the trivial phase). We couple a monochromatic pump at frequency $\nu-\nu_L = 4.6J$ into an edge link ring with strength $\kappa_{\mathrm{ex}} = J/2$, solving its nonlinear propagation equation in the time domain~\cite{supplementary} and including moderate two photon absorption $\kappa_{\mathrm{NL}} = 0.1 \nu_{NL}$ representative of the resonators used in Ref.~\cite{hafezi2013}. At a critical power the pump converges to the stable steady state shown in Fig.~\ref{fig:pump_profile}(a,b), resulting in an average shift to the coupling parameter of $\phi_0-\phi \approx 0.1\pi$, which is sufficient to induce a transition to the Chern insulator phase for the probe field.

In Fig.~\ref{fig:pump_profile}(c), we compute the transmission spectrum of a weak probe beam tuned to the site bands. Without the pump, the site bands are topologically trivial and the band gap forms a deep transmission minimum. When the pump is applied, the mid-gap transmission is increased by $\approx 20$dB due to the formation of an additional resonance associated with a topological edge state. Fig.~\ref{fig:pump_profile}(d) plots the probe's intensity profile, which is directed along the edge and to output port 2 despite the disorder induced by the inhomogeneity of the pump. 

It is interesting to note that the large increase in the transmission of Fig.~\ref{fig:pump_profile}(c) is not easily achievable using the previously studied self-focusing nonlinearity of a bright probe beam. First, two-photon absorption will inevitably result in increased attenuation of a high power probe~\cite{nonlinear_review}. Furthermore, modulational instability may break up the topological edge states~\cite{ablowitz2014,instability,kartashov2016}. On the other hand, using this pump-probe scheme, the nonlinearity is confined to the link rings, where the probe photons spend relatively little time. Consequently, parametric mixing processes between pump and probe are suppressed, and to a good approximation the probe beam dynamics remain governed by a linear Hamiltonian, and the well-established robustness of linear topological edge states holds. For these reasons, pump-induced cross phase modulation is a highly promising alternative for ultrafast and robust nonlinear switching of topological edge states. 

{\it Outlook.---}We have designed a lattice of coupled ring resonators that serves as a promising setting for electro-optic or nonlinearly-reconfigurable topological waveguiding at optical frequencies.  The distinguishing feature of this scheme is the use of a bipartite lattice; for a given circulation, the corresponding tight-binding model is $\mathcal{T}$-broken and has both nearest-neighbor and next-nearest-neighbor couplings, which are of equal magnitude. The phase diagram contains both conventional insulator and Chern insulator phases, and the lattice can be tuned between them via experimentally-accessible parameters. 

Note also that previous methods relying on aperiodic couplings~\cite{hafezi2011,hafezi2013} or dynamic modulation~\cite{fang2012,peano2015,minkov2016} introduce an additional length or energy scale to the system that constrains the bandwidth of the topological phase. For example, dynamic modulation requires the inter-resonator coupling to be much weaker than the modulation frequency; Ref.~\cite{minkov2016} estimated a maximum practical band gap of 33 GHz using terahertz Kerr modulation. Such constraints are bypassed in next-nearest-neighbor coupled models, enabling broadband operation approaching the rings' free spectral range. 

We demonstrated switching of topological edge states based on static detuning of the link resonators or nonlinear cross phase modulation of the site resonators. The latter requires nonlinear resonance shifts of approximately 50 GHz (5\% of the free spectral range), which is well within reach of current silicon photonics technology.  Our scheme can be readily generalized other classes of simple lattices, such as hexagonal or honeycomb lattices, where the larger number of coupled neighbors will result in Chern insulators with longer range coupling, broader bandwidths, and multiple band gaps. 

\acknowledgements{This research was supported by the Institute for Basic Science in Korea (IBS-R024-Y1), the Singapore MOE Academic Research Fund Tier 2 (MOE2015-T2-2-008), the Singapore MOE Academic Research Fund Tier 3 (MOE2016-T3-1-006), AFOSR MURI Grant No. FA95501610323, ONR, the Sloan Foundation, and the Physics Frontier Center at the Joint Quantum Institute.}

\begin{widetext}

\begin{center}
  {\large \textbf{Supplemental Material}}

  for

  {\large Reconfigurable topological phases in next-nearest-neighbor coupled resonator lattices}

  {\footnotesize Daniel Leykam, S. Mittal, M. Hafezi, and Y.~D.~Chong}
\end{center}

\makeatletter 
\renewcommand{\theequation}{S\arabic{equation}}
\makeatother
\setcounter{equation}{0}

\makeatletter 
\renewcommand{\thefigure}{S\@arabic\c@figure}
\makeatother
\setcounter{figure}{0}

\noindent

In this Supplemental Material we derive the tight binding approximation for a periodic ring resonator lattice, verify its accuracy by comparing its band structure to that obtained from the full transfer matrix method, present the tight binding equations for an inhomogeneous lattice, and demonstrate the robustness of the topological edge state-mediated transmission to disorder in the resonant frequencies of the rings.

\section{Scattering matrix model}

Light propagation through resonator lattices is governed by a sequence of evanescent couplings between neighboring rings, and phases accumulated as light propagates between different coupling regions. Each coupling region is described by a scattering matrix,
\be 
\hat{S} (\theta) = \left[ \begin{array}{cc} \cos \theta & -i \sin \theta \\ -i \sin \theta & \cos \theta \end{array} \right],
\ee
parametrized by a coupling angle $\theta$. We have assumed there is negligible backscattering, such that clockwise and counter-clockwise circulating modes are decoupled. Looking for periodic Bloch wave solutions, the optical field amplitude within different parts of a unit cell are determined by a set of 8 amplitudes, $(a_j,b_j,s_j)$, as indicated in Fig.~\ref{fig:ring_schematic}.

\begin{figure}[h]

\includegraphics[width=0.4\columnwidth]{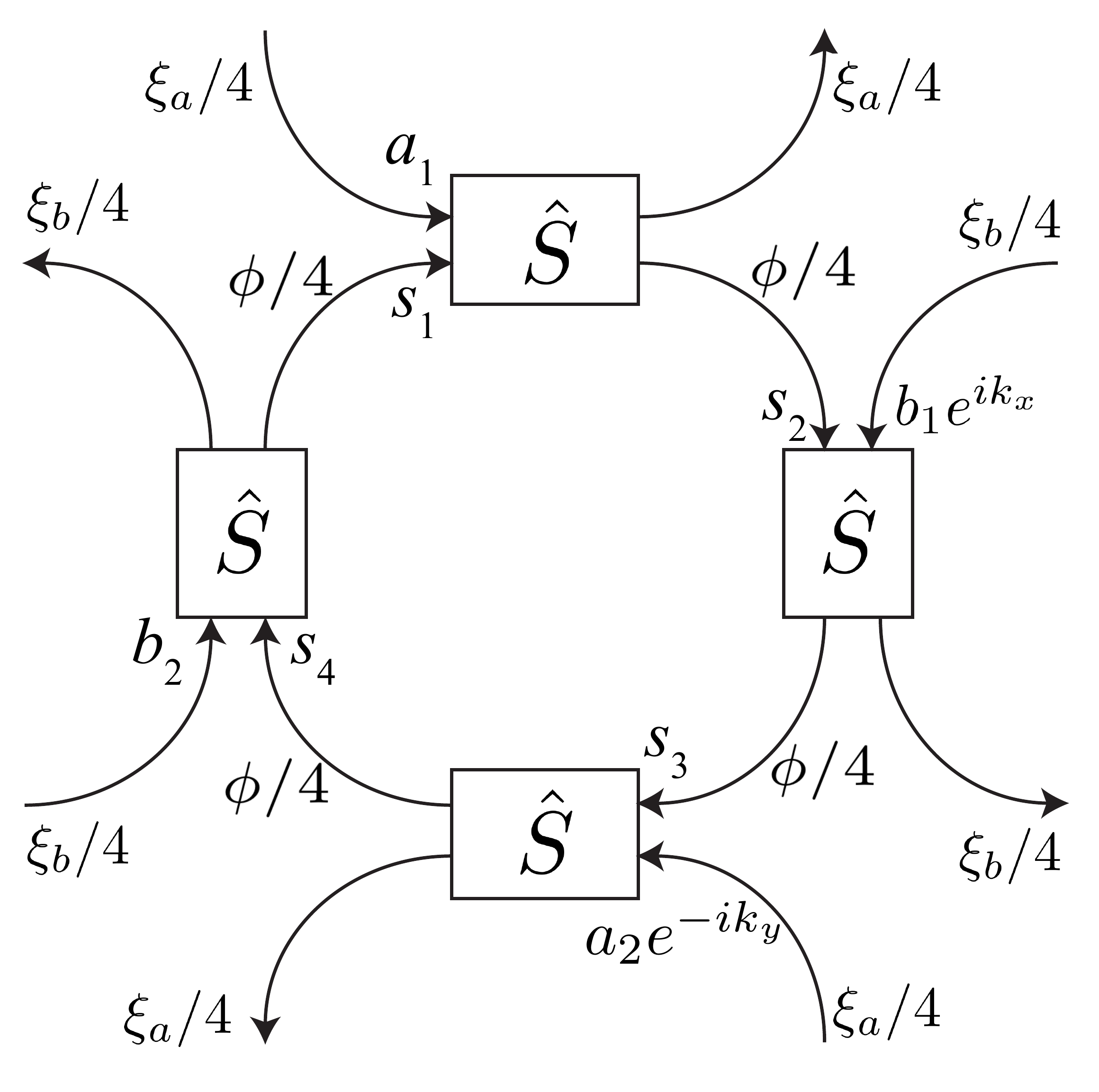}

\caption{Field amplitudes ($a_j, b_j, s_j$) and phase delays ($\xi_j,\phi$) in a unit cell of the resonator lattice.}

\label{fig:ring_schematic}

\end{figure}

The scattering equations describing modes with anti-clockwise circulation in the site rings are
\begin{align}
\left[ \begin{array}{c} s_2 e^{-i \phi/4} \\ a_2 e^{-i \xi_a/2} \end{array} \right] &= \hat{S} \left[ \begin{array}{c} s_1 \\ a_1 \end{array} \right], \label{eq:s1} \\
\left[ \begin{array}{c} s_3 e^{-i \phi/4} \\ b_2 e^{i (k_x - \xi_b/2)} \end{array} \right] &= \hat{S} \left[ \begin{array}{c} s_2 \\ b_1 e^{i k_x} \end{array} \right], \label{eq:s2} \\
\left[ \begin{array}{c} s_4 e^{-i \phi/4} \\ a_1 e^{-i (k_y + \xi_a/2)} \end{array} \right] &= \hat{S} \left[ \begin{array}{c} s_3 \\ a_2 e^{-i k_y} \end{array} \right], \label{eq:s3} \\
\left[ \begin{array}{c} s_1 e^{-i \phi/4} \\ b_1 e^{-i\xi_b/2 } \end{array} \right] &= \hat{S} \left[ \begin{array}{c} s_4 \\ b_2 \end{array} \right],  \label{eq:s4}
\end{align}
where $\xi_j$ and $\phi$ are the round trip phases accumulated in the site and link rings respectively. 

Eqs.~(\ref{eq:s1}-\ref{eq:s4}) can be inverted to obtain the transfer matrix equations
\be 
\left[ \begin{array}{c} s_1 \\ s_4 \end{array} \right] = \hat{T}_1 \left[ \begin{array}{c} b_1 \\ b_2 \end{array} \right], 
\left[ \begin{array}{c} s_2 \\ s_3 \end{array}\right] = \hat{T}_2 \left[ \begin{array}{c} s_1 \\ s_4 \end{array}\right],
e^{i k_x} \left[ \begin{array}{c} b_1 \\ b_2 \end{array} \right]   = \hat{T}_3 \left[ \begin{array}{c} s_2 \\ s_3 \end{array} \right],
\ee
where
\begin{align}
\hat{T}_1 &= i \csc \theta \left[ \begin{array}{cc} e^{i (\phi/4 - \xi_b/2)} \cos \theta & -e^{i \phi/4} \\ e^{-i \xi_b/2} & - \cos \theta \end{array} \right], \quad \hat{T}_3 = -i \csc \theta \left[ \begin{array}{cc} \cos \theta & -e^{-i\phi/4} \\ e^{i \xi_b/2} & - e^{i(\xi_b/2 - \phi/4)} \cos \theta \end{array}\right], \\
\hat{T}_2 &= \frac{1}{\cos\theta (e^{i \xi_a}-1)} \left[ \begin{array}{cc} \frac{1}{2} e^{i \phi/4} (1-\cos 2 \theta + 2 e^{i\xi_a} ) & e^{i ( \xi_a/2 + k_y)} \sin^2 \theta \\ -e^{i (\xi_a/2 - k_y)} \sin^2 \theta & e^{-i \phi/4} ( e^{i \xi_a} \cos^2 \theta - 1) \end{array} \right].
\end{align}
We assume the site and link rings have similar lengths (i.e. approximately the same free spectral range $\mathrm{FSR}=10^3$GHz) and measure frequencies with respect to the site ring resonances, letting $\phi \rightarrow \pi + 2\pi \delta \nu / \mathrm{FSR}$, $\xi_a = 2\pi (\delta\nu - M) / \mathrm{FSR}$, and $\xi_b = 2\pi (\delta\nu + M) / \mathrm{FSR}$. Then given $\delta \nu$ and $k_y$, the eigenvalues of the transfer matrix $\hat{T} = \hat{T}_3 \hat{T}_2 \hat{T}_1$ give the wavenumber $k_x$, allowing reconstruction of the bulk band structure. We use the mean site and link ring intensities $I_a = \sum_j |a_j|^2/2$, $I_b = \sum_j |b_j|^2/2$, and $I_s = \sum_j |s_j|^2/4$ to characterize the distribution of the eigenmodes between the site and link rings, 
\be 
\rho = \frac{I_a+I_b}{I_s + I_a + I_b },
\ee
with $\rho = 1$ (0) corresponding to perfect localization to the sites (links). The cross phase modulation model used in the main text assumes the link ring modes have $\rho \ll 1$. Typical values near the link band edge are $\rho \approx 0.02$ for $J=2$ GHz and $\rho \approx 0.1$ for $J=10$ GHz, indicating the nonlinearity of the sites can be neglected to a first approximation when considering the propagation of a bright pump beam tuned to the link resonance (and vice versa).

\section{Site ring tight binding model}

To obtain the tight binding models we assume weak coupling and small frequency shifts, i.e. $\theta = \sqrt{\frac{4 \pi J}{\mathrm{FSR}}}$ and $4\pi J/\mathrm{FSR} \sim 2\pi (\delta \nu \pm M) / \mathrm{FSR} \sim \epsilon$, with $\epsilon$ a small parameter. We solve the first three scattering equations to obtain the amplitudes $(a_2,b_2,s_j)$, and then solve the final scattering equation to leading order in $\epsilon$ to obtain $b_1/a_1$ and $\delta \nu$,
\begin{align}
\delta \nu &= d_0 \pm |\bs{d}| , \quad |\bs{d}|^2 = d_x^2 + d_y^2 + d_z^2,  \\
d_0 &= J \csc \frac{\phi}{2} (\cos k_x + \cos k_y) + 2J \cot \frac{\phi}{2},\\
d_x &= 2J \csc \frac{\phi}{2} \cos \frac{\phi}{4} (\cos \frac{k_x-k_y}{2} + \cos \frac{k_x+k_y}{2}),\\
d_y &= -2J \csc \frac{\phi}{2} \sin \frac{\phi}{4} (\cos \frac{k_x-k_y}{2} - \cos \frac{k_x+k_y}{2}),\\
d_z &= M + J \csc \frac{\phi}{2} (\cos k_y - \cos k_x), \\
\frac{b_1}{a_1} &= \frac{-d_z \pm \delta \nu}{d_x - i d_y}.
\end{align}
This corresponds to the eigenvalues $\delta\nu$ and eigenvectors $\psi = (a_1,b_1)$ of the Hamiltonian $\hat{H} = d_0 + \bs{d} \cdot \bs{\hat{\sigma}}$ and is equivalent to the site ring tight binding model of Eq.~(2) in the main text.

In Fig.~\ref{fig:spectrum_comparison} we compare the spectra obtained using the transfer matrix and tight binding models for different values of the coupling strength $J$ and link detuning $\Delta \nu = \mathrm{FSR} \phi/(2\pi)$. We obtain good agreement for weakly coupled rings ($J=2$ GHz), while for stronger coupling $J=10$ GHz the tight binding model overestimates position of the upper band edge. Nevertheless, there is still reasonable agreement between the two at lower frequencies in the vicinity of the and gap, justifying our use of the tight binding model in the main text.

\begin{figure}

\includegraphics[width=0.7\columnwidth]{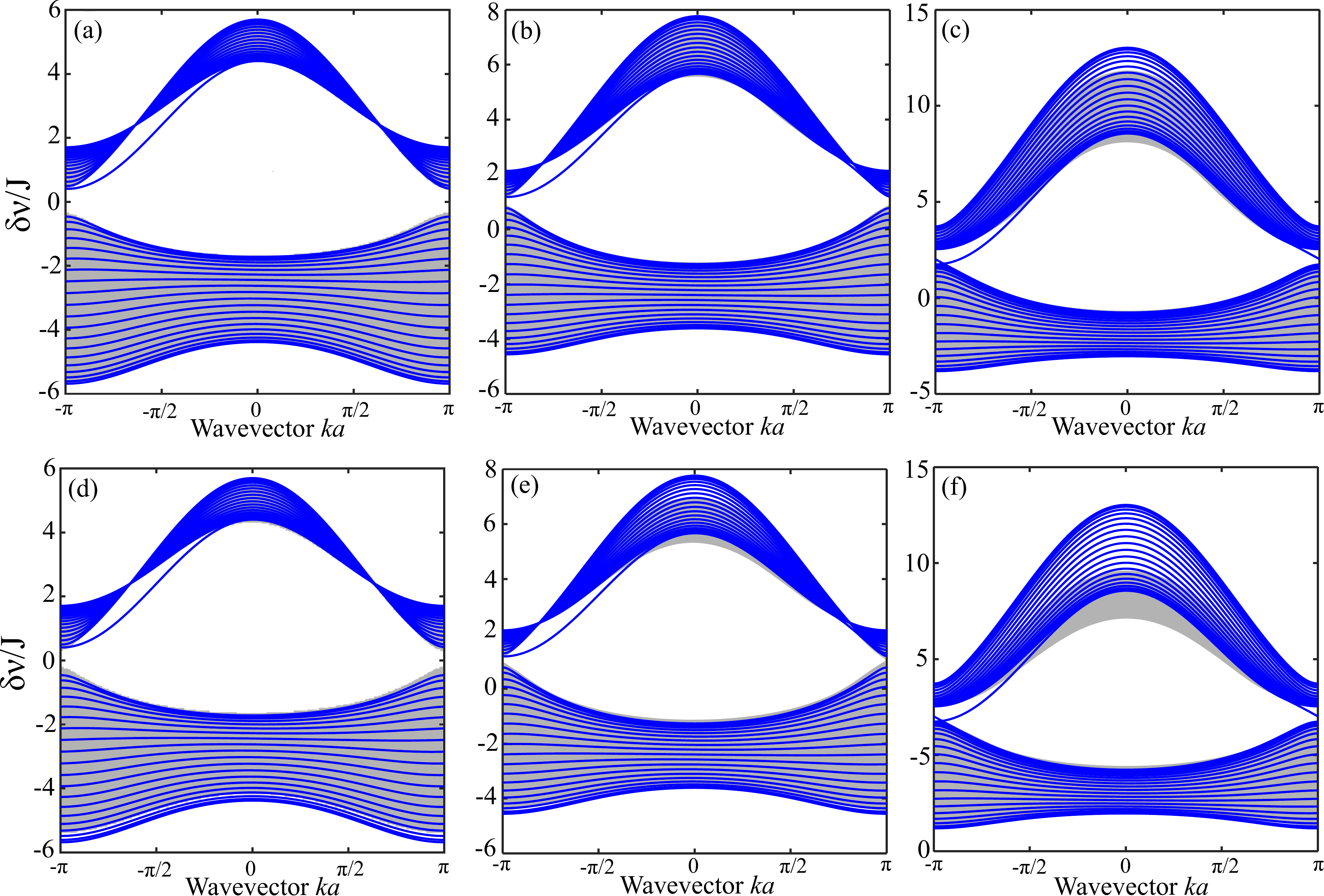}

\caption{Comparison between site ring bands obtained from the transfer matrix method (dark shaded regions) and the tight binding approximation for a semi-infinite lattice of width $N=20$ unit cells with $M=2.4J$. (a,b,c) $J=2$ GHz. (d,e,f) $J=10$ GHz. (a,d) $\phi=\pi$. (b,e) $\phi = 0.7\pi$. (c,f) $\phi = 0.4\pi$.}

\label{fig:spectrum_comparison}

\end{figure}

We illustrate further the behavior of the bulk bands (obtained using the transfer matrix method) as a function of the coupling strength $J$ with Fig.~\ref{fig:spectrum_comparison_2}. For moderate values of the coupling $J$ the positions of the band edges predicted by the transfer matrix and tight binding models differ by less than 10\%. For example, at $J=10$ GHz the relative errors in the inner and outer band edges predicted by the tight binding model are 5\% and 3.5\% respectively (absolute errors $\lesssim 2$ GHz). Interestingly, we note that the tight binding model underestimates the size of the topological band gap. At $J \approx 60$ GHz the site and link bands cross to form a strongly coupled anomalous Floquet topological insulator phase, and the two models begin to give qualitatively different results. Therefore the tight binding model provides a reasonable approximation for $J \lesssim 60$ GHz. 

\begin{figure}

\includegraphics[width=0.3\columnwidth]{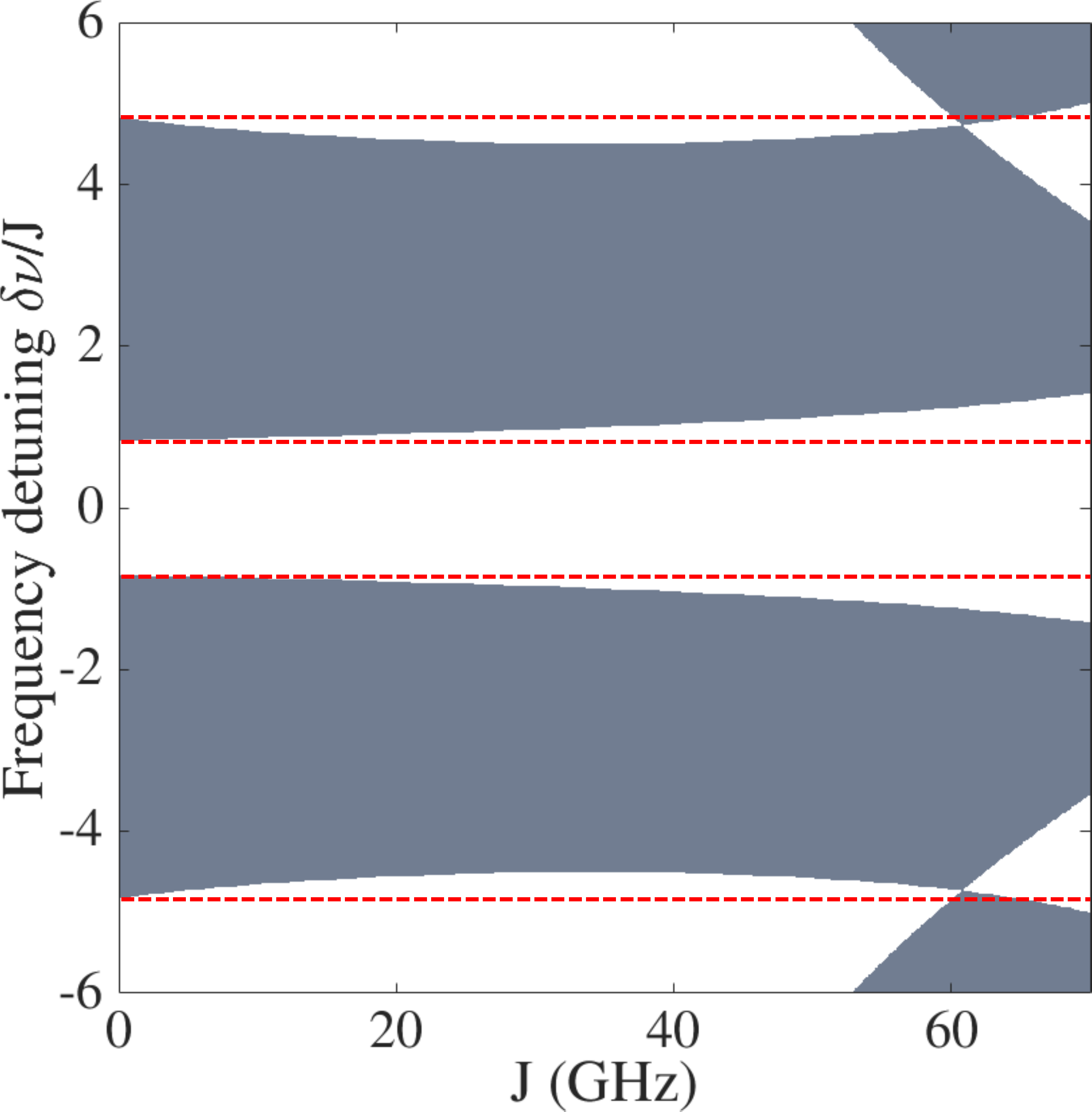}

\caption{Evolution of the bulk spectrum obtained from the transfer matrix method (dark shaded regions) as a function of the coupling strength $J$. The pair of site bands intersect the third link band at $J \approx 60$ GHz, indicating a transition from the Chern insulator phase to an anomalous Floquet insulator phase. For comparison, red dashed lines mark the band edges $\delta \nu / J = \pm 2 (\sqrt{2} \pm 1)$ obtained under the tight binding approximation. Here $M=0$, $\phi = \pi$, and $\mathrm{FSR} = 1000$ GHz.}

\label{fig:spectrum_comparison_2}

\end{figure}

The effective tight binding Hamiltonian for the site rings is given in the Fourier basis, which assumes a perfectly periodic lattice. To obtain equations in real space, we perform an inverse Fourier transform of the Bloch Hamiltonian $\hat{H}$ and then allow its parameters to be position-dependent. The resulting real space tight binding Hamiltonian is
\begin{align}
&\hat{H} = \sum_{x,y} \left(\hat{H}_a + \hat{H}_b + \hat{H}_{ab} + \hat{H}_{ab}^{\dagger} \right), \\
&\hat{H}_a = \hat{a}_{x,y}^{\dagger} \left( \left[ M+W^{(a)}_{x,y}+ J\cot \frac{\phi_{x,y}}{2} + J\cot \frac{\phi_{x,y+1}}{2} \right] \hat{a}_{x,y} + J\csc \frac{\phi_{x,y}}{2} \hat{a}_{x,y-1} + J\csc \frac{\phi_{x,y+1}}{2} \hat{a}_{x,y+1} \right), \\
&\hat{H}_b = \hat{b}_{x,y}^{\dagger} \left( \left[ -M+W^{(b)}_{x,y}+ J\cot \frac{\phi_{x,y}}{2} + J\cot \frac{\phi_{x-1,y}}{2} \right] \hat{b}_{x,y} + J\csc \frac{\phi_{x,y}}{2} \hat{b}_{x+1,y} + J\csc \frac{\phi_{x-1,y}}{2} \hat{b}_{x-1,y} \right), \\
&\hat{H}_{ab} = J \hat{a}_{x,y}^{\dagger} \left( e^{i \phi_{x,y}/4} \hat{b}_{x,y} + e^{i \phi_{x,y+1}/4} \hat{b}_{x+1,y+1} \right) + J \hat{b}_{x,y}^{\dagger} \left( e^{i \phi_{x,y}/4} \hat{a}_{x,y-1} + e^{i \phi_{x-1,y}/4} \hat{a}_{x-1,y} \right),
\end{align}
where $W^{(j)}_{x,y}$ and $\phi_{x,y}$ are now position-dependent site and link ring detunings, with
\be 
\phi_{x,y} = \phi_0 - 2 \pi (W_{x,y}^{(s)} + 2\nu_{\mathrm{NL}} |\psi^{(s)}_{x,y}|^2)/\mathrm{FSR},
\ee
accounting for linear and nonlinear detunings of the link resonances induced by the pump beam $s_{x,y}$. In experimental implementations these equations account for the most significant source of disorder (imperfections in the ring resonance frequencies), with the hopping disorder being less important. 

To estimate the impact of disorder on the topological edge state-mediated transmission in Fig.~2 of the main text, we take take $W^{(j)}_{x,y}$ to be normally-distributed random variables with standard deviation $0.8J$ and solve the linear scattering problem for 1000 random realizations of the disorder. The disorder-averaged spectra in Fig.~\ref{fig:domain_disorder} show that the transmission in the topological band gap is indeed robust and only weakly affected by this moderately-strong disorder; while the edge state transport is ballistic (in the absence of losses), in finite systems the disorder can reduce the average transmission by lowering the effective coupling strength to the external leads. This effect can be mitigated by coupling the leads to many lattice sizes to minimize the influence of individual site detunings.

\begin{figure}

\includegraphics[width=0.6\columnwidth]{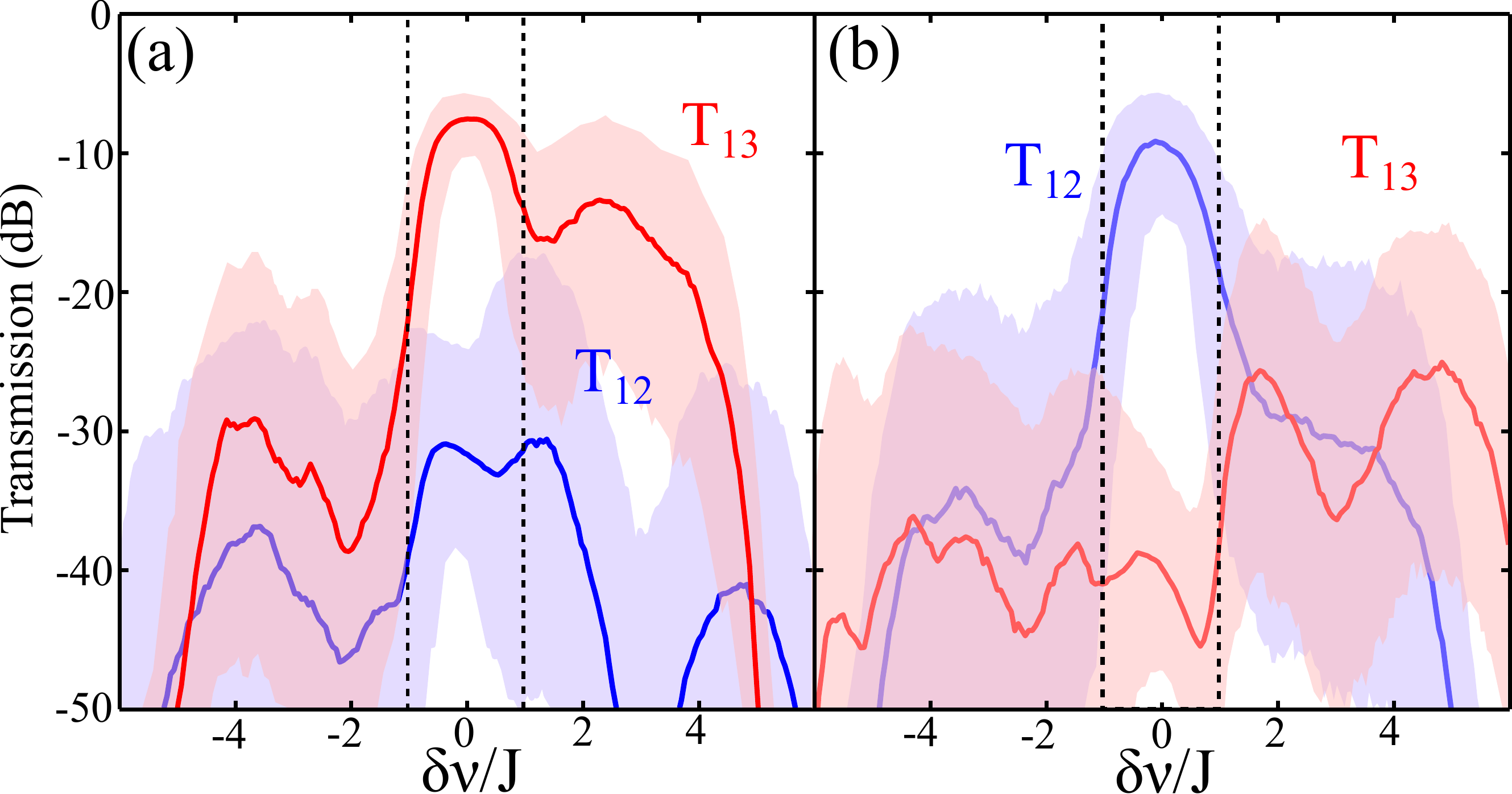}

\caption{Transmission through the domain walls of Fig.~2 of the main text in the presence of Gaussian disorder in the ring resonances with standard deviation $0.8J$. Solid lines indicate average values, dashed lines the band gap, and shaded regions are 90\% confidence intervals estimated from 1000 realizations of the disorder. }

\label{fig:domain_disorder}

\end{figure}

We also show in Fig.~\ref{fig:scaling} how the disordered transmission scales with the system size. The contrast between the edge and bulk transmission grows with the system size, due to the onset of Anderson localization of the bulk modes, which can be seen by inspecting the corresponding field profiles. Therefore ideal topologically-protected transport is recovered for large systems, with disorder localizing the bulk states leading to vanishing transmission, while the edge states avoid Anderson localization.

\begin{figure}

\includegraphics[width=0.6\columnwidth]{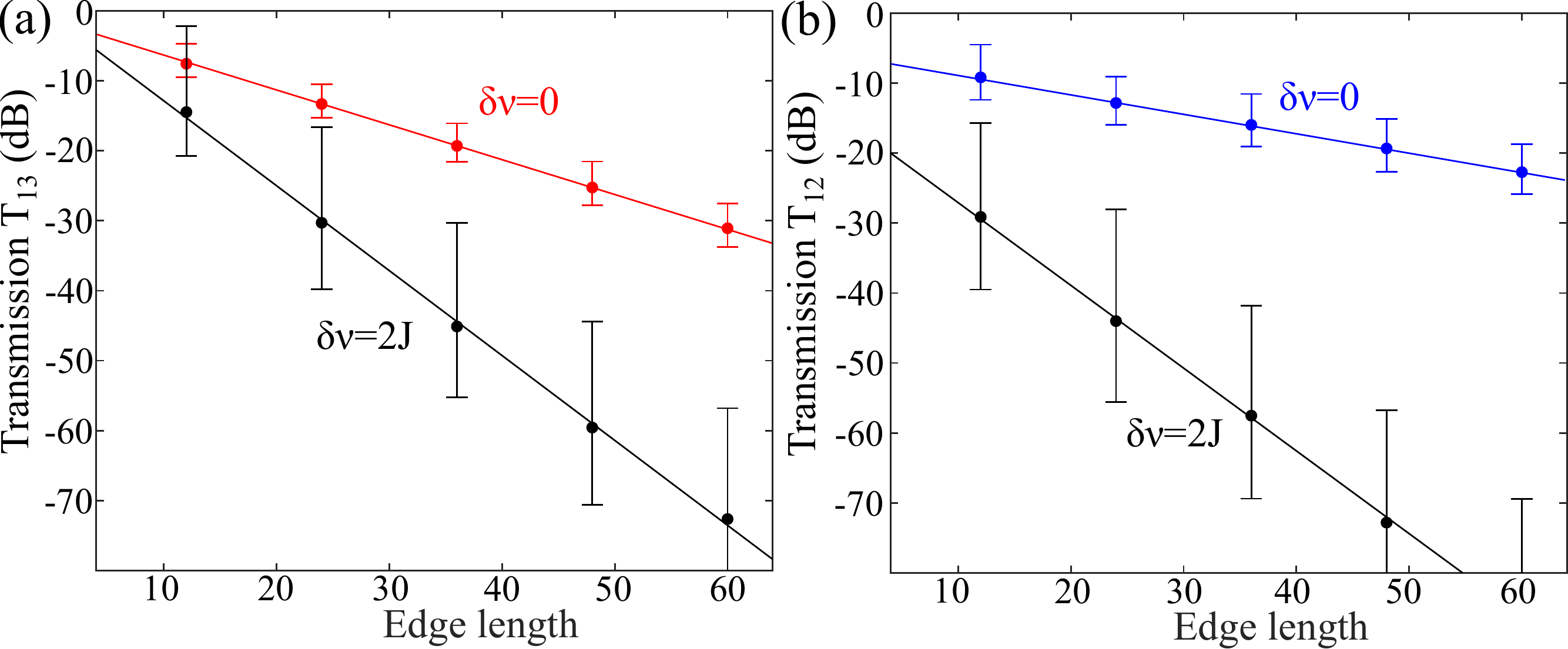}

\caption{Scaling of the average transmission in Fig.~\ref{fig:domain_disorder} with the system size, taking representative probe frequencies in the band gap ($\delta \nu = 0$) and in one of the bulk bands ($\delta \nu = 2J$) with the system size. Error bars indicate 90\% confidence intervals. The edge length $3N/2$ is the number of edge sites between the input and output ports, where the overall system size is $N\times N$ unit cells.}

\label{fig:scaling}

\end{figure}

Note that we have only considered the case of disorder in the ring resonance frequencies, which is the most significant source of disorder in typical experiments. Since our system is analogous to a quantum spin Hall insulator, the disorder-robust transmission only holds as long as the two pseudo-spins remain decoupled. Surface roughness of the rings can couple the pseudo-spins and induce backscattering. However, previous experiments have observed that this class of disorder is negligible compared to imperfections in the ring resonance frequencies; for example, the measured backscattered signal in Ref.~[21] was about 40dB weaker than the transmitted signal. This is similarly the case for 1D CROW devices, which are limited by Anderson localization induced by random site frequencies, rather than backscattering due to surface roughness. Therefore our model provides topological protection against the most significant class of disorder affecting coupled resonator lattices.

\section{Link ring tight binding model}

The tight binding Hamiltonian governing the pump beam $\psi^{(s)}_{x,y}$ is similarly obtained by considering small detunings from the link resonance, $\delta \nu_L = \delta \nu - \frac{\phi_0 \mathrm{FSR}}{2\pi}$ and again solving the scattering equations Eq.~(\ref{eq:s1}-\ref{eq:s4}) perturbatively for small frequency shifts, yielding the spectrum
\be 
\delta \nu_L = -2J \csc \frac{\phi_0}{2} ( \cos k_x + \cos k_y ) - 4 J \cot \frac{\phi_0}{2},
\ee
which corresponds to propagation dynamics governed by a real space tight binding Hamiltonian on a square lattice,
\be 
\hat{H}_s =\sum_{x,y} \left[ \delta \nu_L \hat{s}_{x,y}^{\dagger} \hat{s}_{x,y} + \hat{s}_{x,y}^{\dagger} J \csc \frac{\phi_0}{2} \left( \hat{s}_{x,y-1} + \hat{s}_{x,y+1} + \hat{s}_{x-1,y} + \hat{s}_{x+1,y} \right)\right], \label{eq:site_model}
\ee
where $\phi_0 = 2\pi\Delta \nu / \mathrm{FSR}$ is determined by the detuning between the site and link rings $\Delta \nu$. The small detuning $\delta \nu_L$ from the single link resonance $\nu_L$
\be 
\delta \nu_L = \nu - \nu_L - i \kappa + (\nu_{\mathrm{NL}} - i \kappa_{\mathrm{NL}} ) |\psi^{(s)}_{x,y}|^2  + 4 J \cot \frac{\phi_0}{2}, \label{eq:site_resonance}
\ee
includes intrinsic losses $\kappa$, cubic focusing nonlinearity $\nu_{\mathrm{NL}}$, and two photon absorption $\kappa_{\mathrm{NL}}$. Due to the high symmetry of the square lattice formed by the link rings, the same tight binding equations govern both choices of the mode handedness; there are no synthetic gauge fields and the Bloch wave eigenmodes of Eq.~\eqref{eq:site_model}, $\psi^{(s)}_{x,y} = \exp[ i (k_x x + k_y y)]$, form a single trivial band. More complex lattice geometries such as the honeycomb similarly do not exhibit nontrivial phases in this weak coupling limit because of this effective time reversal symmetry.

From this approximate solution of the scattering equations one also finds that the intensities in the site rings are suppressed by a factor of $2\pi \csc^2 \frac{\phi_0}{2} J/\mathrm{FSR}$ compared to the link ring intensities, which is consistent with the numerical transfer matrix solution above. Notice also that the small detunings $\pm M$ to the sites do not affect the coupling seen by the link rings (to leading order in $J/\mathrm{FSR}$). Similarly, the dynamics of the pump beam will be insensitive to small nonlinearity-induced detunings to the site rings, justifying the assumption of linear site rings.

\end{widetext}

\begin{thebibliography}{99}

\bibitem{topological_photonics_review}
L. Lu, J.~D. Joannopoulos, and M.~Solja\u{c}i\'{c}, {\it Topological photonics}, Nature Photon. {\bf 8}, 821 (2014).

\bibitem{review_2}
Y. Wu, C. Li, X. Hu, Y. Ao, Y. Zhao, and Q. Gong, {\it Applications of topological photonics in integrated photonic devices}, Adv. Opt. Mat. 1700357 (2017).

\bibitem{ma2015}
T. Ma, A.~B. Khanikaev, S.~H. Mousavi, and G. Shvets, {\it Guiding electromagnetic waves around sharp corners: topologically protected photonic transport in metawaveguides}, Phys. Rev. Lett. {\bf 114}, 127401 (2015).

\bibitem{cheng2016}
X. Cheng, C. Jouvaud, X. Ni, S. H. Mousavi, A.Z. Genack, and A.~B. Khanikaev, {\it Robust reconfigurable electromagnetic pathways within a photonic topological insulator}, Nature Materials {\bf 15}, 542 (2016).

\bibitem{goryachev2016}
M. Goryachev and M.~E. Tobar, {\it Reconfigurable microwave photonic topological insulator}, Phys. Rev. App. {\bf 6}, 064006 (2016).

\bibitem{susstrunk2017}
R. S\"usstrunk, P. Zimmermann, and S.~D. Huber, {\it Switchable topological phonon channels}, New J. Phys. {\bf 19}, 015013 (2017).

\bibitem{lumer2013}
Y. Lumer, Y. Plotnik, M.~C. Rechtsman, and M. Segev, {\it Self-localized states in photonic topological insulators}, Phys. Rev. Lett. {\bf 111}, 243905 (2013).

\bibitem{ablowitz2014}
M.~J.~Ablowitz, C.~W.~Curtis, and Y.-P.~Ma, {\it Linear and nonlinear traveling edge waves in optical honeycomb lattices}, Phys.~Rev.~A {\bf 90}, 023813 (2014).

\bibitem{nonlinear_SSH}
Y.~Hadad, A.~B.~Khanikaev, and A.~Alu, {\it Self-induced topological transitions and edge states supported by nonlinear staggered potentials}, Phys.~Rev.~B {\bf 93}, 155112 (2016).

\bibitem{leykam2016}
D. Leykam and Y.~D. Chong, {\it Edge solitons in nonlinear-photonic topological insulators}, Phys. Rev. Lett. {\bf 117}, 143901 (2016).

\bibitem{instability}
Y. Lumer, M.~C. Rechtsman, Y. Plotnik, M. Segev, {\it Instability of bosonic topological edge states in the presence of interactions}, Phys. Rev. A {\bf 94}, 021801(R) (2016).
  
\bibitem{kartashov2016}
Y.~V. Kartashov and D. Skyrabin, {\it Modulational instability and solitary waves in polariton topological insulators}, Optica {\bf 3}, 1228 (2016).

\bibitem{hadad2017}
Y. Hadad, V. Vitelli, and A. Alu, {\it Solitons and propagating domain walls in topological resonator arrays}, ACS Photon. {\bf 4}, 1974 (2017).

\bibitem{NJP_paper}
X. Zhou, Y. Wang, D. Leykam, and Y.~D. Chong, {\it Optical isolation with nonlinear topological photonics}, New J. Phys. {\bf 19}, 095002 (2017).

\bibitem{shalaev2017}
M.~I. Shalaev, S. Desnavi, W. Walasik, N.~M. Litchinitser, {\it Reconfigurable topological photonic crystal}, New J. Phys. {\bf 20}, 023040 (2018).


\bibitem{rechtsman2013}
M.~C. Rechtsman, J.~M. Zeuner, Y. Plotnik, Y. Lumer, D. Podolsky, F. Dreisow, S. Nolte, M. Segev, and A. Szameit, {\it Photonic Floquet topological insulators}, Nature {\bf 496}, 196 (2013).

\bibitem{leykam2016a}
  D.~Leykam, M.~C.~Rechtsman, and Y.~D.~Chong, {\it Anomalous topological phases and unpaired Dirac cones in photonic Floquet topological insulators}, Phys.~Rev.~Lett.~{\bf 117}, 013902 (2016).

\bibitem{noh2017}
J. Noh, S. Huang, D. Leykam, Y. D. Chong, K. Chen, and M. C. Rechtsman, {\it Experimental observation of optical Weyl points and Fermi arc-like surface states}, Nature Phys. {\bf 13}, 611 (2017).

\bibitem{hafezi2011}
  M.~Hafezi, E.~A.~Demler, M.~D.~Lukin, and J.~M.~Taylor, {\it Robust
    optical delay lines with topological protection}, Nature~Phys.~\textbf{7}, 907 (2011).

\bibitem{fang2012}
K. Fang, Z. Yu, and S. Fan, {\it Realizing effective magnetic field for photons by controlling the phase of dynamic modulation}, Nature Photon. {\bf 6}, 782 (2012).


\bibitem{hafezi2013}
M. Hafezi, S. Mittal, J. Fan, A. Migdall, and J.~M. Taylor, {\it Imaging topological edge states in silicon photonics}, Nature Photon. {\bf 7}, 1001 (2013).

\bibitem{mittal2014}
S. Mittal, J. Fan, S. Faez, A. Migdall, J.~M. Taylor, and M. Hafezi, {\it Topologically robust transport of photons in a synthetic gauge field}, Phys. Rev. Lett. {\bf 113}, 087403 (2014).

\bibitem{peano2015}
V. Peano, C. Brendel, M. Schmidt, and F. Marquardt, {\it Topological phases of sound and light}, Phys. Rev. X {\bf 5}, 031011 (2015).

\bibitem{minkov2016}
M. Minkov and V. Savona, {\it Haldane quantum Hall effect for light in a dynamically modulated array of resonators}, Optica {\bf 3}, 200 (2016).


\bibitem{mittal2016b}
S. Mittal, S. Ganeshan, J. Fan, A. Vaezi, and M. Hafezi, {\it Measurement of topological invariants in a 2D photonic system}, Nature Photon. {\bf 10}, 180 (2016).

\bibitem{electro_modulator}
C. Qiu, W. Gao, R. Vajtai, P.~M. Ajayan, J. Kono, and Q. Xu, {\it Efficient modulation of 1.55$\mu$m radiation with gated graphene on a silicon microring resonator}, Nano Lett. {\bf 14}, 6811 (2014).

\bibitem{nonlinear_review}
J. Leuthold, C. Koos, and W. Freude, {\it Nonlinear silicon photonics}, Nature Photon. {\bf 4}, 535 (2010).


\bibitem{haldane_model}
F.~D.~M. Haldane, {\it Model for a quantum Hall effect without Landau levels: condensed-matter realization of the ``parity anomaly''}, Phys. Rev. Lett. {\bf 61}, 2015 (1988).

\bibitem{liang2013}
  G.~Q.~Liang and Y.~D.~Chong, {\it Optical resonator analog of a two-dimensional topological insulator},
  Phys.~Rev.~Lett.~\textbf{110}, 203904 (2013).

\bibitem{pasek2014}
  M.~Pasek and Y.~D.~Chong, {\it Network models of photonic Floquet topological insulators}, Phys.~Rev.~B {\bf 89}, 075113 (2014).
  
\bibitem{hu2015}
  W.~Hu, J.~C.~Pillay, K.~Wu, M.~Pasek, P.~P.~Shum, and Y.~D.~Chong,
  {\it Measurement of a topological edge invariant in a microwave
    network}, Phys.~Rev.~X \textbf{5}, 011012 (2015).
  
\bibitem{spoof_plasmon}
F. Gao, Z. Gao, X. Shi, Z. Yang, X. Lin, H. Xu, J. D. Joannopoulos, M. Soljacic, H. Chen, L. Lu, Y. Chong, and B. Zhang, {\it Probing topological protection using a designer surface plasmon structure}, Nature Commun. {\bf 7}, 11619 (2016).  
  
\bibitem{shi2017}
T. Shi, H.~J. Kimble, and J.~I. Cirac, {\it Topological phenomena in classical optical networks}, Proc. Natl. Acad. Sci. USA {\bf 114}, E8967 (2017).



\bibitem{NNN_hopping}
W. Beugeling, J.~C. Everts, and C.~M. Smith, {\it Topological phase transitions driven by next-nearest-neighbor hopping in two-dimensional lattices}, Phys. Rev. B {\bf 86}, 195129 (2012).


\bibitem{supplementary}
See online supplemental material.

\bibitem{agarwal_book}
G. Agarwal, {\it Nonlinear fiber optics} (Academic, 2007).

\end{thebibliography}
\end{document}